\documentclass{lmcs}
\pdfoutput=1

\usepackage{lastpage}
\lmcsdoi{22}{2}{14}
\lmcsheading{}{\pageref{LastPage}}{}{}%
{Jul.~03,~2025}{May~06,~2026}{}


\usepackage{mathtools}
\usepackage{amsmath}
\usepackage{amssymb}
\usepackage{amsthm}
\usepackage{graphicx} 
\usepackage{hyperref}
\usepackage[utf8]{inputenc}

\newtheorem{problem}[thm]{Problem}

\newcommand\bits{\{0,1\}}
\newcommand\uu{{\bits^*}}

\newcommand\mm{{\bits^m}}

\newcommand\tpv{{T_{\mbox{\tiny PV}}}}
\newcommand\pp{\mbox{P}}
\newcommand\np{\mbox{NP}} 
\newcommand\npco{{\np \cap \conp}}
\newcommand\conp{{\mbox{co}\np}}
\newcommand\ddp{\mbox{DD}_P}
\newcommand\dd{\dot {\bigvee} }
\newcommand\lpv{{L_{\mbox{\tiny PV}}}}
\newcommand\spv{{S^1_2(\mbox{PV}})}

\newcommand\tr{|\mkern-2.5mu|}
\newcommand\sat{\mbox{SAT}}
\newcommand\taut{\mbox{TAUT}}

\begin{document}

\title{On $\npco$ proof complexity generators}  

\author[J.~Kraj\'{\i}\v{c}ek]{Jan Kraj\'{\i}\v{c}ek\lmcsorcid{0000-0003-0670-3957}} 

\address{Faculty 
of Mathematics and Physics, Charles University, Prague}
\email{jan.krajicek@protonmail.com}

\begin{abstract}
	\noindent
	Motivated by the theory of proof complexity generators
we consider the following $\Sigma^p_2$ search problem $\ddp$ determined by a 
propositional proof system $P$: given
a $P$-proof $\pi$ of a disjunction $\bigvee_i \alpha_i$, no two $\alpha_i$ having an atom in common, find $i$ such that $\alpha_i \in \mbox{TAUT}$.

We formulate a hypothesis (ST) that for some strong proof system $P$ the problem
$\ddp$ is not solvable
in the student-teacher model with a p-time student and a constant number of rounds.
The hypothesis follows from the existence of hard one-way permutations.

We prove, using a model-theoretic assumption, that (ST) implies $\np \neq co\np$.
The assumption concerns the existence of extensions of models of a bounded arithmetic theory
and it is open at present if it holds.  
\end{abstract}

\maketitle

\section*{Introduction}

Proof complexity generators are some maps $g : \uu \rightarrow \uu$ 
with $|g(x)|$ determined by
and larger than $|x|$. Their purpose is to "generate" hard tautologies: propositional formulas
$\tau(g)_b$ expressing that $b \notin Rng(g)$. To be able to express this propositionally
we need that $g$ is p-time or more generally $\npco$ (possibly non-uniform).

For $\npco$ maps\footnote{These are maps for which the binary predicate {\em the $i$-th bit of $g(x)$} is in $\npco$.} 
$g$ sending $n$ bits to $m > n$ bits and $b \in \mm$,
the statement $b \notin Rng(g)$ has the form 
\begin{equation} \label{fo}
\forall x (|x| = n) \exists i < m \forall y_i\ A_b(x,i,y_i)
\end{equation}
with $A_b$ an open formula in the language $\lpv$ having a function symbol for 
every p-time clocked
algorithm, and expressing that $y_i$ does not witness that the $i$-th bit of $g(x)$ is $b_i$, the $i$-th bit of $b$.

The provability of such a statement in the true universal $\lpv$-theory $\tpv$ 
can be analyzed using the interactive student-teacher model (abbr. S-T) of \cite{KPS}. 
In this model S (the student) upon receiving $x := a$ proposes its first candidate solution
$i_1$. If it is correct T (the teacher) will acknowledge this and the computation stops
with $i_1$ as its output. Otherwise she will send to S a counter-example, some $w_1$ such that
$\neg A_b(a,i_1,w_1)$. In that case the computation enters the second round with S proposing a second candidate solution $i_2$ and T either accepting it or returning a counter-example
$w_2$. The computation goes for a number of rounds until a solution is found.

We are interested in the case where the number of rounds is bounded by a constant. The S-T model
makes sense for general total search problems where 
the condition for the acceptance of a solution,
here $\forall y\ \neg A_b(x,i,y)$, starts with a universal quantifier. Here all such search problems will be $\Sigma^p_2$ and following \cite{k4} we denote by
$$
\mbox{ST}[\mbox{F}, t(n)]
$$
the class of 
total $\Sigma^p_2$ search problems such that there is a student whose moves are computed
by an algorithm from $\mbox{F}$ that solves the problems in $t(n)$ rounds
interacting with any teacher. 

There are two candidate $\npco$ proof complexity generators: an NW generator and a gadget
generator based on a non-deterministic circuit as the gadget, cf. \cite{k4}. 
An analysis using the 
S-T model for (\ref{fo}) for the NW-generator was performed in \cite{Kra-nwg}
and it was shown that (\ref{fo}) cannot be witnessed in $\mbox{ST}[\pp/poly, O(1)]$
if a one-way permutation hard for p-size circuits exists. The gadget generator was analyzed
in \cite{k4} and a similar result was shown assuming a certain plausibly looking
computational hardness hypothesis (K), cf. \cite[Sec.8.6]{k4}. 

These first-order constructions gave independence results for the theory $\tpv$.
There is a close relation between first-order theories and (tacitly propositional) 
proof systems, based on the notion of propositional translations (for theories 
in $\lpv$ this goes back to \cite{Coo75,KP-jsl}). 
It is also a classic fact that lengths-of-proofs lower bounds for formulas $\gamma$ 
are essentially equivalent to 
constructing models of a suitable theory where $\gamma$ can be falsified.

When trying to get from the independence results for the two $\npco$ generators
lengths-of-proofs lower bounds for the corresponding $\tau$-formulas one runs into the issue that the propositional translation is not entirely faithful in the following sense
(this is discussed in detail as a "missing reflection" in \cite{Kra-nwg}).
 
The $\tau(g)$ formulas for an $\npco$-map $g$ have the form
\begin{equation} \label{dp}
\gamma(p,q) \ :=\ \bigvee_i \beta_i(p, q^i)	
\end{equation}
with tuples of atoms $p, q = (q^i)_i$ and tuples $q^i$ disjoint. 
A falsifying assignment to $\gamma$ exists in a model iff the model satisfies
\begin{equation}\label{fla1}
\exists x, y \forall i < m \ \neg A_b(x,i, (y)_i)
\end{equation}
where $y$ is a list of strings $(y)_i$, $i < m$. 
The constructions in \cite{Kra-nwg,k4} yield models satisfying the negation of 
(\ref{fo})
\begin{equation}\label{fla2}
\exists x \forall i < m \exists y_i	\ \neg A_b(x,i, y_i)\ .
\end{equation}
To get (\ref{fla1}) from (\ref{fla2})
one needs the sharply bounded
collection scheme BB of \cite{Bus-book} for open $\lpv$-formulas allowing to switch
the two inner quantifiers. Unfortunately
by \cite{CT} this principle is not provable in $\tpv$ (unless factoring is not hard)
while adding it to $\tpv$ makes the proof-theoretic analysis of (\ref{fo}) from
\cite{Kra-nwg,k4} invalid (the KPT theorem of \cite{KPT} does not apply in any usable form
to this non-universal theory).

In this note we propose a model-theoretic method how to bypass the BB scheme
and get lengths-of-proofs lower bounds from a hypothesis that a certain $\Sigma^p_2$ search problem is hard for S-T computations. The model-theoretic assumption is of the form 
saying that certain extensions of models of $\tpv$ exist. The caveat is that such 
extensions are known to exist for models satisfying the BB scheme for open formulas
but it is not known in general. However, model theory seems to offer much more room
to vary arguments than proof theory does and we think it is worthwhile to try to 
assert the model theoretic statement needed. 

\bigskip

The paper is organized as follows. The relevant search problem and the hypothesis about its
S-T intractability are discussed in Section \ref{2}, the model theoretic assumption 
is explained in Section \ref{3} and the theorem reducing 
lengths-of-proofs lower bounds to Hypothesis (ST)
is stated and proved in Section \ref{4}.

\smallskip

We assume that the reader is familiar with basic proof complexity theory.
We do use quite a lot of notions and facts and it is unfeasible to review it here
(the review would be substantially longer than the technical part). However,
the reader can find all what we use in \cite{prf}. We formulate the result and the arguments without referring to proof complexity generators to make it available also to readers not familiar with that theory, although that is the
primary motivation for this research. The reader can find 
everything that we use in \cite{k4}.

A convention:
letters $p,q,\dots$, possibly with superscripts $p^i, q^i, \dots$ denote {\em tuples} 
of propositional atoms.

\section{The search problem and its S-T computability} \label{2}

A propositional disjunction $\bigvee_{i < r} \alpha_i$ is a {\bf disjoint disjunction}, 
denoted
$$
\dd_{i < m} \alpha_i\ ,
$$ 
if no two $\alpha_i$ have an atom in common. We often leave $m$ out, 
and write just $\dd_i \alpha_i$, as $m$ is bounded by the size of the formula. 
Recall from \cite{CooRec} that a
propositional proof system (abbr. pps) is a p-time decidable binary relation
$P(x,y)$ such that the condition $\exists x P(x,y)$ defines $\taut$. 
Instead of writing $P(\sigma,\beta)$ we
use the suggestive notation $\sigma : P \vdash \beta$ for the provability relation:
$\sigma$ is a $P$-proof of $\beta$. The symbol $P \vdash \beta$ means that $\beta$ has
a $P$-proof.

Given a pps $P$, 
the {\bf $\Sigma^p_2$ search problem $\ddp$} is:

\begin{itemize}

\item {\em input}: a pair $\pi, \dd_i \alpha_i$ s.t. $\pi : P \vdash \dd_i \alpha_i$,
	
\item {\em solution}: any $i$ s.t. $\alpha_i \in \taut$.
\end{itemize}
The problem can be solved in the S-T model described in the Introduction.
S, upon receiving an input, proposes his first candidate solution $i_1$. 
If it is correct T acknowledges it. Otherwise she sends to S a counter-example:
an assignment falsifying $\alpha_{i_1}$. S then proposes his second candidate solution, using also the counter-example he learned in the first round, and the computation proceeds
analogously for some number of rounds.

Following \cite[Def.2.4.3]{k4} we define a {\bf strong pps} to be an
extension of EF
(Extended Frege, cf. \cite{CooRec}) by a p-time set of tautologies as extra axioms (see \cite [Thm.2.4.4]{k4} for their nice properties). Every pps can be p-simulated by a strong one.

\bigskip
\noindent
{\bf Hypothesis (ST):} {\em There exists a strong pps $P$ such that
$\mbox{DD}_P \notin \mbox{ST}[\pp, O(1)]$, $\pp$ standing here also
for the class of p-time algorithms 
computing functions.
}

\smallskip

In fact, we do think that the hypothesis is true for all strong proof systems 
(which is equivalent to saying that it holds for EF, the Extended Frege proof system)
but for the statement this apparently weaker formulation suffices.

\medskip

The $\mbox{DD}_P$ problem is a special case of a problem with a larger class of inputs
applying to formulas of the form (\ref{dp});
we shall call the problem here $\mbox{D}_P$:
\begin{itemize}
	
\item {\em input}: a triple $\pi, \bigvee_i \beta_i(p,q^i), a$ 
s.t. $\pi : P \vdash \bigvee_i \beta_i$, only atoms $p$ can occur in more than one $\beta_i$, and $a$ is a truth assignment to atoms $p$,
	
\item {\em solution}: any $i$ s.t. $\beta_i(a, q^i) \in \mbox{TAUT}$.
\end{itemize}

\begin{lem}
$$
\mbox{DD}_{P} \notin \mbox{ST}[\pp, O(1)] \Leftrightarrow
\mbox{D}_{P} \notin \mbox{ST}[\pp, O(1)]
$$
\end{lem}

The possibility that $\mbox{D}_{P} \in \mbox{ST}[\pp, O(1)]$ holds for strong proof systems 
was considered in \cite{PS}.
However, it was shown in \cite{Kra-limitations} that this is not true (assuming the existence
of strong one-way permutations), essentially for reasons similar to why the feasible interpolation fails for them (cf. \cite[Chpt.18]{prf}).
The problem to witness (\ref{fo}) studied in \cite{Kra-nwg} does not have a $P$-proof as a
part of the 
input and this led to the need to consider non-uniform p-time students there.
Another plausibly looking hypothesis about the feasible intractability of a certain search problem related to non-deterministic circuits and implying (ST) was given in \cite{k4}
(cf. hypothesis (K) there).

We state for the record the plausibility of (ST).

\begin{lemC} [\cite{Kra-limitations}]
(ST) is true if a strong one-way permutation exists (no polynomial size circuit can invert
it with a non-negligible advantage).	
\end{lemC}

\section{Bounded arithmetic models} \label{3}

The classic theorem we shall recall now uses the correspondence between
first-order theories and proof systems mentioned in the Introduction.
A prototype of the statement was proved in \cite[Thm.1]{KP-model} working with 
theory $S^1_2$ of \cite{Bus-book} and its relation to Extended Frege proof
system EF (a consequence of \cite{Coo75}).
The same argument applies to any pair of corresponding theories and proof systems
using the generalization of \cite{Coo75} developed in \cite{KP-jsl}.
The underlying argument is discussed in detail also in \cite[Sec.20.1]{prf}.

As we are aiming at all proof systems at once we can work with the theory $\tpv$ and avoid
discussing the correspondence between theories and proof systems. The first item of the theorem speaks about propositional proof systems and by those we mean proof systems determined by standard p-time binary relations and not proof systems in the sense of the non-standard model.
In particular, a pps $P$ may be incomplete in a model of $\tpv$ but it will always be sound as that is expressible by a universal $\lpv$-formula.

\begin{thm} [after {\cite[Thm.1]{KP-model}}] \label{kp}
Assume $\mathbf M$ is a non-standard model of $\tpv$ and let $\varphi \in {\mathbf M}$
be a propositional formula (in the sense of the model). The following two statements are equivalent:

\begin{enumerate}

\item $\varphi$ has no proof in $\mathbf M$ in any proof system.

\item There exists an extension ${\mathbf M}' \supseteq {\mathbf M}$ such that
$$
{\mathbf M}' \ \models \ \tpv\ +\ (\neg \varphi) \in \sat\ .
$$	
\end{enumerate}
\end{thm}

The fact that the construction in \cite{KP-model} works with models of $S^1_2$
implies that ${\mathbf M}'$ preserves (i.e. will satisfy)
all $\Sigma^b_1$ properties of elements of $\mathbf M$ true there
(this uses that the models satisfy the sharply bounded collection principle BB
for open $\Sigma^b_1$-formulas, a consequence of $S^1_2$ (cf. \cite{Bus-book,kniha}).

Further it is remarked at the end of \cite{KP-model} that one can arrange that ${\mathbf M}'$ introduces no new lengths
into ${\mathbf M}$: 
\begin{equation} \label{equ}
	Log({\mathbf M}') = Log({\mathbf M})
\end{equation}
where $Log({\mathbf M})\ =\ \{|m|\ |\ m \in {\mathbf M}\}$. 
A construction which yields this (with an additional requirement on $\mathbf M$)
is the forcing construction underlying \cite[Thm.9.4.2]{kniha}.

The problem we are interested in
is whether these two properties, that is item 2 of Theorem \ref{kp} and (\ref{equ}), can be arranged too
if we only know that ${\mathbf M} \models \tpv$.
We shall formulate the problem in a less direct way but closer to the construction 
in the proof of \cite[Thm.9.4.2]{kniha}. This formulation also relates better
to the possibility to use Boolean-valued models discussed at 
the end of \cite{Kra-pseudofinite}.

\begin{problem} \label{problem}
Assume $\mathbf M\models \tpv$ and that it satisfies Condition 1 in Theorem \ref{kp}.
Does it follow that there are $\lpv$-structures ${\mathbf M}^*, {\mathbf M}'$:
$$
{\mathbf M} \subseteq {\mathbf M}^* \subseteq {\mathbf M}'
$$
such that

\begin{enumerate}

\item ${\mathbf M}^*$ preserves all $\Sigma^b_1(PV)$-properties of elements
of ${\mathbf M}$,

\item ${\mathbf M}' \models \tpv + (\neg \varphi) \in\sat$,

\item  $Log({\mathbf M}^*) = Log({\mathbf M}')$?
\end{enumerate}
\end{problem}
The assumption we shall use in Theorem \ref{main}
is that the problem has the affirmative solution. Note that
the negative answer implies $\pp \neq \np$ as otherwise\footnote{If $f$ is a p-time algorithm
	that finds a satisfying assignment for all satisfiable formulas then this fact, being a universal statement, would be in $\tpv$ and hence $\tpv$ would prove all true bounded formulas.} 
$\tpv$ proves $\spv$.

\medskip

The problem asks about all models of $\tpv$ but it would suffice to get the affirmative answer
for a sufficiently universal class of models (e.g. countable or recursively saturated).
Also, it suffices if ${\mathbf M}^*$ preserves from $\mathbf M$ properties of the form 
$\forall i \exists y B(x,i,y)$ with $i$ sharply bounded, $y$ bounded and $B$ open.

\medskip

Let us remark that constructing extensions of 
bounded arithmetic models introducing no new lengths
is very close to constructing expansions of non-standard finite structures. Such expansions
are at the heart of applications of model theory to lower bounds for circuits or proofs
as it is spectacularly demonstrated in \cite{Ajt83,Ajt88}. 
Developing such constructions is in my view a crucial task in proof
complexity; the survey \cite{Kra-pseudofinite} discusses this in detail (cf. also \cite[Chpt.20]{prf})).

\section{The theorem} \label{4}

Universal $\lpv$ sentences $\forall x A(x)$ can be translated into a sequence of p-size 
propositional formulas $\tr A\tr^n$ such that 
$$
\forall x (|x|=n)A(x) \leftrightarrow \tr A\tr^n \in \taut\ 
$$ 
(cf. \cite{kniha} or \cite[Sec.12.6]{prf}).

For a pps $P$ let $Ref_P$ be the universal valid $\lpv$ formula
$$
x : P \vdash y\ \rightarrow (\neg y) \notin \sat
$$
where we simplify the notation and write $\neg y$ for the p-time function sending a code
of a formula to the code of its negation.

\begin{thm} \label{main}
Assume that Problem \ref{problem} has the affirmative solution. Then Hypothesis (ST) 
implies $\np \neq co\np$.
\end{thm}

\begin{proof}\phantom{x}

\medskip\noindent\textbf{(1)}

Assume for the sake of a contradiction that the hypothesis (ST) and 
$\np = \conp$ both hold. We shall bring it to a contradiction using the model-theoretic
assumption. 
Take a strong pps $P$ that satisfies the condition in (ST) and is also p-bounded.

\medskip\noindent\textbf{(2)}

Consider theory $S$ in the language extending $\lpv$ by two constant symbols: $\pi$ and
$\alpha$, and having the following axioms:
\begin{itemize}

\item $\tpv$,

\item $\alpha$ is a disjoint disjunction of the form
 $\dd_{i<m} \alpha_i$ ($m$ is determined by $\alpha$),
  
\item $\pi : P \vdash \alpha$,	

\item $\forall i<m\ (\neg\alpha_i) \in \sat$.	
\end{itemize}
If $S$ were inconsistent then 
applying the KPT theorem (cf. \cite{KPT,kniha, k4}) to the universal theory
with axioms listed in the first three items would give p-time functions 
computing moves of a student
that solves $\ddp$ in a constant number of rounds, thus violating (ST).

Let $\mathbf M$ be a model (necessarily non-standard) of $S$.

\medskip\noindent\textbf{(3)}

Assume $c \geq 1$ is a constant such that any $\beta \in \taut$ has a $P$-proof of size
$\le |\beta|^c$. We shall abbreviate by
$$
\sigma : P \vdash_* \beta
$$
the formula 
$$
(\sigma : P \vdash \beta)\ \wedge \ (|\sigma|\le |\beta|^c)\ .
$$
In the following formula the
symbols $(y)_i$ and $(z)_i$ denote the $i$-th string in the list coded by $y, z$, respectively, and $len(y) \le |y|$ denotes the number of these strings.

\medskip

\noindent{\bf Claim:} {\em 
For any pps $Q$ the universal formula $A_Q$:
\begin{equation} \label{claim}
x : P\vdash \dd_i (y)_i \ \rightarrow\ 
Q \not\vdash \tr \forall i < len(y)\ (z)_i : P \not\vdash_* (y)_i	\tr 
\end{equation}
(we leave the obvious length-bounds in the translation out)
is true and hence in $\tpv$.
}

\smallskip

The claim holds because some 
$(y)_i$ has to be a tautology and hence - by the p-boundedness of $P$ - would 
have a short $P$-proof.

\medskip\noindent\textbf{(4)}

All $A_Q$ hold in $\mathbf M$ and we can substitute $x := \pi$ and $y := \alpha$.
This yields
$$
{\mathbf M} \ \models\ 
Q \not \vdash \varphi(r) 
$$
where $\varphi(r)$ is obtained from
$$
\tr \forall i < len(y)\ (z)_i : P \not\vdash_* (y)_i	\tr (r,q)
$$
by substituting the bits of $\alpha$ for atoms $q$ (hence 
the sharply bounded universal quantifier becomes $\bigwedge_{i < m}$)
and the remaining atoms $r$ correspond to (bits of)
$z$.

\medskip\noindent\textbf{(5)}

Now we conclude the proof by using the assumption that Problem \ref{problem} has the affirmative
solution. This gives us ${\mathbf M} \subseteq {\mathbf M}^*\subseteq {\mathbf M}'$
such that
\begin{equation}\label{log}
Log({\mathbf M}^*) = Log({\mathbf M}')\ ,
\end{equation}
and ${\mathbf M}'\models \tpv \wedge \neg\varphi\in \sat$.
Note that ${\mathbf M}^*$ satisfies the last axiom of $S$:
\begin{equation} \label{lastax}
{\mathbf M}^*\ \models\ \forall i<m\ (\neg\alpha_i) \in \sat\ 
\end{equation}
as it preserves the $\Sigma^b_1(PV)$-properties from $\mathbf M$.

Let $\sigma \in {\mathbf M}'$ be an assignment to the atoms $r$ that 
falsifies the formula $\varphi$ in ${\mathbf M}'$. 
Hence $\sigma = (\sigma_i)_{i < m}$ and we have
$$
{\mathbf M}' \ \models\ \sigma_{i_0}: P \vdash \alpha_{i_0}
$$
for some $i_0 < m$. But by (\ref{log}) $i_o$ is also already 
in ${\mathbf M}^*$ and hence by (\ref{lastax}) we get:
$$
{\mathbf M}'\ \models\ (\neg \alpha_{i_0}) \in \sat\ .
$$
That contradicts $Ref_P$ which holds in ${\mathbf M}'$ as it models $\tpv$.
\end{proof}

\medskip

Note that the proof yields a somewhat stronger conclusion that in no strong $P$ 
can a proof of some disjunct in a disjoint disjunction be polynomially bounded in size
in terms of a proof of the disjunction (i.e. the strong feasible disjunction property fails
for strong proof systems, cf.\cite{k4})

\section*{Acknowledgment}
\noindent I thank N.Arteche (Lund), O.Je\v zil (Prague) and
M.M\" uller (Passau) for comments on a draft of the paper.


\begin{thebibliography}{KPT91}

\bibitem[Ajt83]{Ajt83}
M.~Ajtai.
\newblock $\sigma^1_1$ - formulas on finite structures.
\newblock {\em Annals of Pure and Applied Logic}, 24:1--48, 1983.

\bibitem[Ajt88]{Ajt88}
M.~Ajtai.
\newblock The complexity of the pigeonhole principle.
\newblock In {\em Proc. IEEE 29$^{\mbox{th}}$ Annual Symp. on Foundation of
  Computer Science}, pages 346--355, 1988.

\bibitem[Bus86]{Bus-book}
S.~R. Buss.
\newblock {\em Bounded Arithmetic}.
\newblock Bibliopolis, Naples, 1986.

\bibitem[cek95]{kniha}
J.~Kraj\'{\i}\v cek.
\newblock {\em Bounded arithmetic, propositional logic, and complexity theory},
  volume~60 of {\em Encyclopedia of Mathematics and Its Applications}.
\newblock Cambridge University Press, 1995.

\bibitem[Coo75]{Coo75}
S.~A. Cook.
\newblock Feasibly constructive proofs and the propositional calculus.
\newblock In {\em Proc. 7$^{\mbox{th}}$ Annual ACM Symp. on Theory of
  Computing}, pages 83--97, 1975.

\bibitem[CR79]{CooRec}
S.~A. Cook and R.~A. Reckhow.
\newblock The relative efficiency of propositional proof systems.
\newblock {\em J. Symbolic Logic}, 44(1):36--50, 1979.

\bibitem[CT06]{CT}
S.~A. Cook and N.~Thapen.
\newblock The strength of replacement in weak arithmetic.
\newblock {\em ACM Transactions on Computational Logic}, 7(4), 2006.

\bibitem[Ka90]{KP-model}
J.~Kraj\'{\i}\v{c}ek and P.~Pudl\' ak.
\newblock Propositional provability in models of weak arithmetic.
\newblock In {\em Computer Science Logic (Kaiserlautern, Oct. '89)}, volume 440
  of {\em Lecture Notes in Computer Science}, pages 193--210, 1990.
\newblock eds. E. Boerger, H. Kleine-Bunning and M. M. Richter.

\bibitem[KP89]{KP-jsl}
J.~Kraj\'{\i}\v{c}ek and P.~Pudl\'{a}k.
\newblock Propositional proof systems, the consistency of first-order theories
  and the complexity of computations.
\newblock {\em J. Symbolic Logic}, 54(3):1063--1079, 1989.

\bibitem[KPS90]{KPS}
J.~Kraj\'\i\v{c}ek, P.~Pudl\'ak, and J.~Sgall.
\newblock Interactive computations of optimal solutions.
\newblock In {\em Mathematical Foundations of Computer Science (B. Bystrica,
  August '90)}, volume 452 of {\em Lecture Notes in Computer Science}, pages
  48--60, 1990.
\newblock B. Rovan (ed.).

\bibitem[KPT91]{KPT}
J.~Kraj\'\i\v{c}ek, P.~Pudl\'ak, and G.~Takeuti.
\newblock Bounded arithmetic and the polynomial hierarchy.
\newblock {\em Annals of Pure and Applied Logic}, 52:143--153, 1991.

\bibitem[Kra11]{Kra-nwg}
J.~Kraj\'{\i}\v{c}ek.
\newblock On the proof complexity of the nisan-wigderson generator based on a
  hard $np \cap conp$ function.
\newblock {\em J. of Mathematical Logic}, 11(1):11--27, 2011.

\bibitem[Kra16]{Kra-pseudofinite}
J.~Kraj\'{\i}\v{c}ek.
\newblock Expansions of pseudofinite structures and circuit and proof
  complexity.
\newblock In {\em Liber Amicorum Alberti}, volume~30 of {\em Tributes Ser.},
  pages 195--203. College Publications, London, 2016.
\newblock eds. Jan van Eijck, Rosalie Iemhoff and Joost J. Joosten.

\bibitem[Kra19]{prf}
J.~Kraj\'{\i}\v{c}ek.
\newblock {\em Proof complexity}, volume 170 of {\em Encyclopedia of
  Mathematics and Its Applications}.
\newblock Cambridge University Press, 2019.

\bibitem[Kra20]{Kra-limitations}
J.~Kraj\'{\i}\v{c}ek.
\newblock A limitation on the kpt interpolation.
\newblock {\em Logical Methods in Computer Science}, 16(3), 2020.

\bibitem[Kra25]{k4}
J.~Kraj\'{\i}\v{c}ek.
\newblock {\em Proof complexity generators}, volume 497 of {\em London
  Mathematical Society Lecture Note Series}.
\newblock Cambridge University Press, 2025.

\bibitem[PS21]{PS}
J.~Pich and R.~Santhanam.
\newblock Strong co-nondeterministic lower bounds for np cannot be proved
  feasibly.
\newblock In {\em Proc. of the 53rd Annual ACM SIGACT Symposium on Theory of
  Computing}, pages 223--233, 2021.

\end{thebibliography}
\end{document}